# CoRuTiGe: A Possible Spin Gapless Semiconductor

Ravinder Kumar[1], Tufan Roy[2,3], Baisali Ghadai[4], Rakesh Kumar[4], Sucheta Mondal[4], Anil Kumar[5], Archana Lakhani[6], Devendra Kumar[6], Masafumi Shirai[2,3], and Sachin Gupta[1,*]

[1]Department of Physics, Bennett University, Greater Noida 201310, India

[2]Center for Science and Innovation in Spintronics, Core Research Cluster, Tohoku University, Sendai 980-8577, Japan

[3]Research Institute of Electrical Communication, Tohoku University, Sendai 980-8577, Japan

[4]Department of Physics, Shiv Nadar Institution of Eminence, Greater Noida 201314, India

[5]Department of Physics, Chaudhary Charan Singh University, Meerut-250004 India

[6]UGC-DAE Consortium for Scientific Research, University Campus, Khandwa Road, Indore 452001, India

*Corresponding author email: sachin.gupta@bennett.edu.in

## Abstract

We report experimental and theoretical investigations on the quaternary Heusler alloy CoRuTiGe, synthesized using the arc melting technique. Crystal structure analysis reveals a tetragonal structure at room temperature. Magnetization measurements as a function of temperature and magnetic field indicate ferromagnetic nature with a saturation magnetization of ~0.681 $\mu_B$/f.u. at 5 K. The temperature dependence of electrical resistivity shows a nearly linear decrease in the high-temperature range, indicating the spin gapless semiconductor (SGS)–like behavior of the material. This SGS nature is further supported by the temperature-independent carrier concentration and mobility. Hall effect analysis reveals that the anomalous Hall effect in CoRuTiGe arises from both intrinsic and extrinsic mechanisms. Additionally, a well-defined symmetric negative magnetoresistance is observed at low temperatures. These findings suggest that CoRuTiGe holds significant promise for spintronic applications.

## I. Introduction

Spin gapless semiconductors (SGSs) are a new class of quantum materials characterized by a unique electronic band structure, where one spin sub-band exhibits semiconducting behavior, while in the other, the valence and conduction bands touch at the Fermi level, reflecting the gapless nature, as shown in Fig. 1 [1–4]. These materials can exhibit quadratic or linear dispersion between energy and momentum [1,2]. Such unique band structures give rise to novel physical properties, making SGSs highly attractive for spintronic applications. The unique band topology of SGSs leads to 100% spin-polarization of both electrons and holes, which is essential for highly efficient spin injection and filtering in devices. Since no energy gap exists for one spin channel, charge carriers can be excited from the valence band to the conduction band with no energy, resulting in extremely low-energy spin transport. In SGSs with linear (Dirac-like) dispersion, the carriers exhibit negligible effective mass, which significantly enhances their mobility compared to conventional semiconductors. Unlike half-metallic ferromagnets (HMFs), where one spin sub-band exhibits semiconducting/insulating behavior and the other remains metallic, the band structure of spin gapless semiconductors is highly sensitive to external factors such as strain, doping, or electric and magnetic fields. This sensitivity arises because in SGSs, the valence and conduction bands in one spin channel touch at the Fermi level, placing the system at a critical point of a zero bandgap [5]. This tunability allows precise control over spin polarization and transport properties, broadening the potential applications of these materials. The capability of controlling spin orientation in the conduction as well as valence bands makes SGSs an ideal candidate for quantum information processing, development of spin-based qubits, quantum Hall effect-based devices, magnetic tunnel junctions and spintronic memory devices [6–9].

Co doped $PbPdO_2$ was first predicted to be SGS by X. L. Wang and later realized experimentally [1,10]. The first experimental demonstration of SGS behavior from Heusler family was reported in an inverse Heusler compound $Mn_2CoAl$ by S. Ouardi *et al* [11]. SGS materials have been observed to exhibit nearly temperature-independent resistivity or conductivity, characterized by an extremely small temperature coefficient of resistivity (TCR), almost constant carrier concentration across temperatures, and a negligible Seebeck coefficient [11]. Several Heusler alloys have been theoretically predicted to be SGSs; however, only a few have been experimentally confirmed. To fully exploit the potential of this novel class of materials, it is essential to identify new materials and verify their properties experimentally.

In this work, we synthesize a quaternary Heusler alloy CoRuTiGe using arc melt technique. We study structural, magnetic, and magneto-transport properties in detail. The material crystallizes in a tetragonal structure and shows ferromagnetic behavior. The temperature and field dependence of electrical resistivity suggest SGS behavior in CoRuTiGe. The experimental results are also supported by theoretical studies.

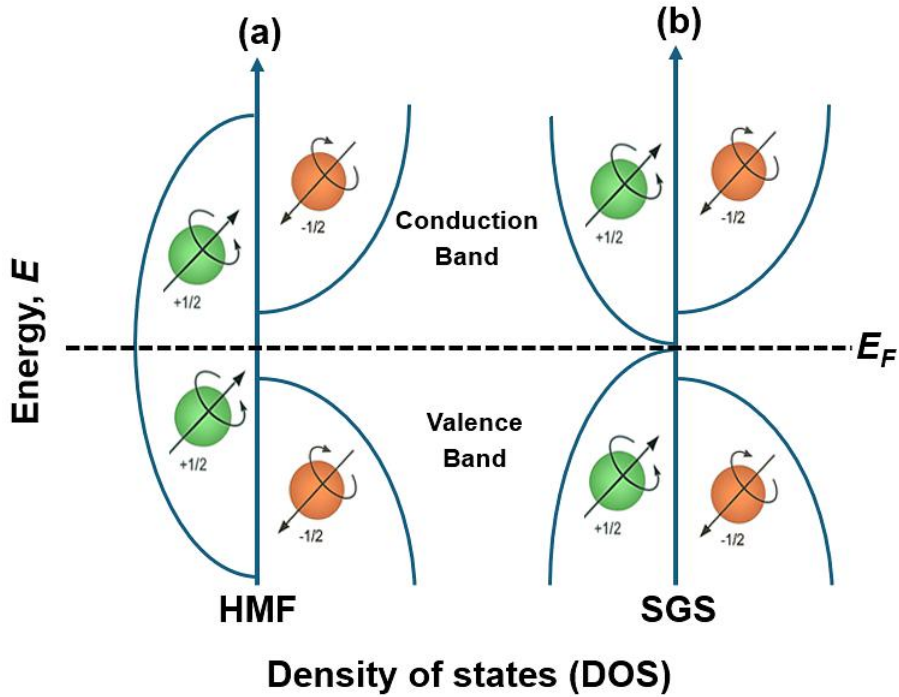

**FIG.** 1. Density of states (DOS) representation for (a) half metallic ferromagnet (HMF), and (b) spin gapless semiconductor (SGS).

## II.  Experimental and theoretical methods

Polycrystalline CoRuTiGe was synthesized under an argon atmosphere using the arc-melt technique. High-purity (at least 99.99 %) constituent elements (Co, Ru, Ti, and Ge) were weighed in a stoichiometric ratio of 1:1:1:1 and melted in a water-cooled copper hearth. The sample was melted 4-5 times by flipping it upside down for better homogeneity of the sample. To improve the crystal structure, the arc-melted sample was sealed in a quartz tube under high vacuum (up to $10^{-5}$ mbar) and annealed in a furnace at 850 °C for 7 days. The sample was subsequently quenched in ice water. The crystal structure of the annealed sample was examined using X-ray diffraction (XRD) analysis, performed with Cu Kα radiation ($\lambda = 1.54$ Å) using Bruker D8 Advance diffractometer at room temperature. Magnetization measurements were carried out using a vibrating sample magnetometer (VSM) associated with Physical Property Measurement System (PPMS), DynaCool, Quantum Design, USA. Longitudinal electrical resistivity was measured using the four-probe technique with an applied current of 5 mA using the PPMS. Hall and magnetoresistance measurements were performed as a function of temperature and magnetic field by applying a current of 20 mA in PPMS.

First-principles based electronic structure calculations have been performed using projector-augmented-wave method as implemented in Vienna Ab-initio Simulation Package (VASP) [12–14]. Generalized gradient approximation (GGA) has been used for the exchange correlation energy/potential [15]. A $k$-mesh of $12\times 12 \times 12$ and a planewave cutoff energy of 520 eV were used for the self-consistent-field calculations. The effect of antisite disorder was studied by the consideration of $2 \times 2 \times 2$ supercell, which includes 32 atoms. The randomness of the swapping disorder was considered by special quasirandom structures (SQS) as implemented in Alloy Theoretic Automated Toolkit (ATAT) package [16,17]. Furthermore, we evaluate Heisenberg exchange coupling constants using the spin-polarized relativistic Korringa-Kohn-Rostoker package (SPR-KKR) [18]. Here, we used scalar relativistic potential, which ignores spin-orbit-coupling (SOC). Thereafter, we use mean-field-approximation to evaluate Curie temperature from the Heisenberg exchange coupling constants using Liechtenstein's formalism [19]. We investigate Berry curvature driven anomalous Hall conductivity (AHC) after constructing a tight binding Hamiltonian with maximally localized Wannier functions. We used the WANNIER90 code and WANNIERTOOLS for the evaluation of AHC [20–22]. $s$, $p$, $d$ orbitals of Co, Ru, and Ti, and $s$, $p$, orbitals of Ge were used to construct Wannier functions. It should be noted that the computational cost associated with larger supercells in disordered structures restricts our investigation of AHC only to the ordered phase.

## III. Results and Discussions

### A. Crystal structure

Fig. 2 shows the room temperature XRD pattern in the $2\theta$ range of 20- 90 degrees along with Rietveld refinement. The bottom plot in Fig. 2 shows the difference of experimentally obtained and calculated data. The Rietveld refinement analysis using FullProf suite [23] reveals tetragonal crystal structure with a space group P42/nnm (space group no. 134). The lattice parameters obtained from Rietveld refinement were found to be $a = b = 6.09$ Å and $c = 5.84$ Å. The reason for transition from the cubic to tetragonal phase could be the stress/strain along the $c$-axis [24]. Various Heusler alloys including $Mn_{3-x}Co_xGa$, $Mn_{3-x}Fe_xGa$ and CoRuTiSn previously have been reported to crystallize in tetragonal structure [25–27].

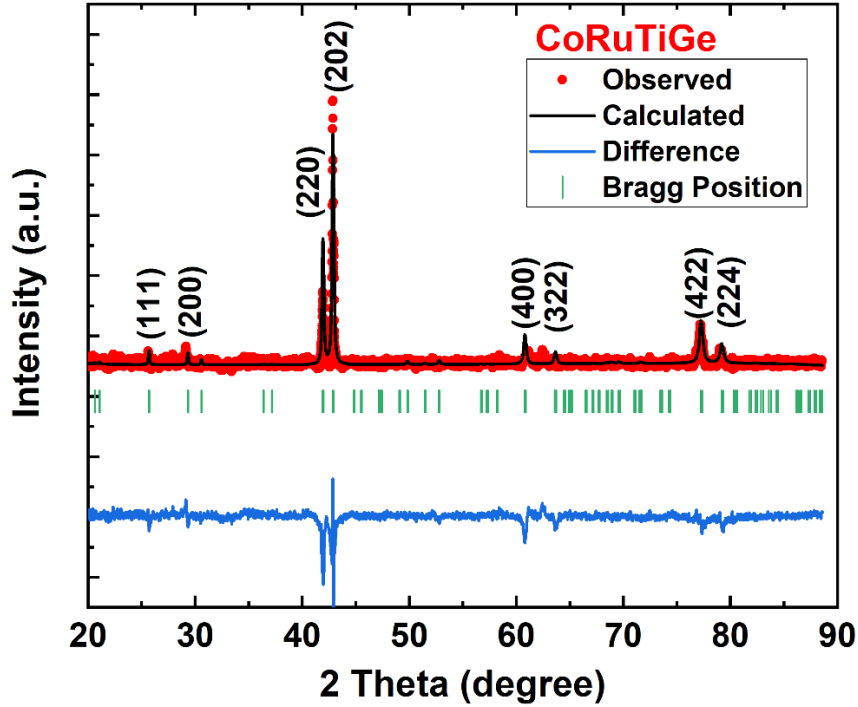

**FIG.** 2. Room temperature powder X-ray diffraction (XRD) pattern along with Rietveld refinement for CoRuTiGe.

**B. Magnetic properties**

Fig. 3(a) shows the temperature, $T$ dependence of magnetization, $M$, under zero-field-cooled (ZFC) and field-cooled (FC) conditions with an applied field of $H = 500$ Oe for CoRuTiGe. The material exhibits small thermomagnetic irreversibility, along with a slight upturn in magnetization at low temperatures. Such behavior is often associated with magnetic disorder, where competing ferromagnetic and antiferromagnetic interactions, possibly arising from site disorder or antistites defects [28]. The magnetic transition from ferromagnetic to paramagnetic phase is observed near 250 K. Fig. 3(b) shows the field-dependent magnetization ($M$−$H$) curves at various temperatures (2, 5, 100, 200, 300, and 400 K). The sample exhibits negligible hysteresis, indicating soft ferromagnetic nature of the material. Total magnetic moment of a quaternary Heusler alloy can be predicted by the Slater-Pauling (S-P) rule using the following equation: [29,30]

$$M_t = (N_V - 24) \; \mu_B/f.u. \qquad (1)$$

Where, $M_t$ is the total magnetic moment and $N_V$ is the total number of valence electrons. CoRuTiGe has 25 valence electrons, therefore the magnetic moment value is anticipated to be 1 $\mu_B$/f.u. However, the observed value of magnetic moment is found to be ~ 0.68 $\mu_B$/f.u, deviating from the theoretical prediction by Slater–Pauling rule (1 $\mu_B$/f.u) [31]. The deviation in the magnetic moment can be attributed to several factors. Most notably, the material crystallizes in a tetragonal structure with a slight deviation in the *c*-lattice parameter, which alters the electronic band structure and weakens spin polarization. Additionally, site disorder, particularly between Co, Ru, and Ti atoms, may induce antiferromagnetic interactions or partial moment compensation, reducing the net magnetic moment. Such disorder also supports the observed FC–ZFC magnetic irreversibility.

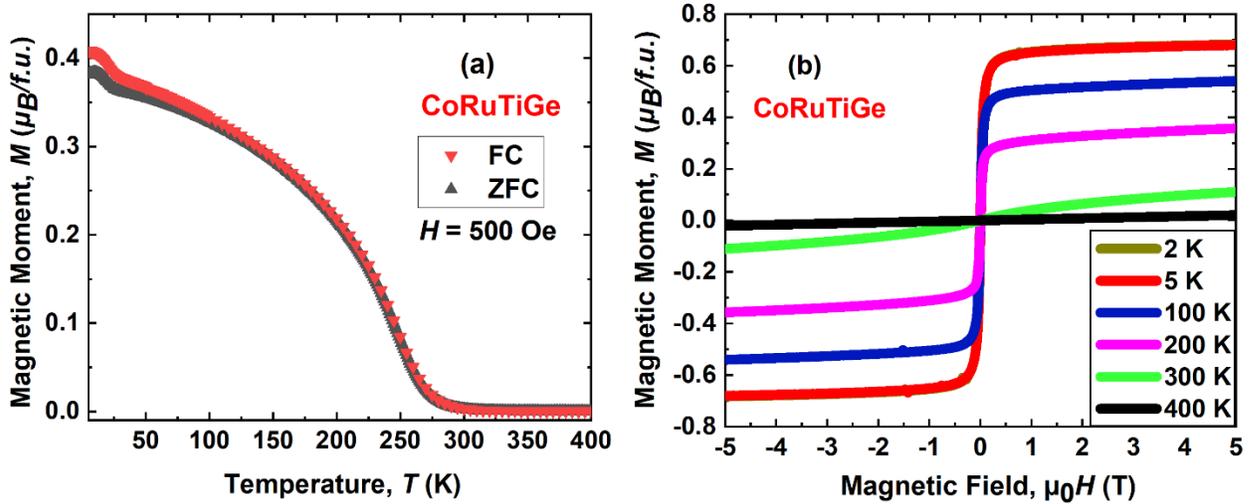

**FIG.** 3. (a) Temperature, *T* dependence of magnetization, *M* in zero field cooling (ZFC) and field cooling (FC) configurations in an applied magnetic field, *H* = 500 Oe. (b) The field, *H* dependence of magnetization, *M* as a function of temperature for CoRuTiGe.

### C. Transport properties

The temperature dependence of electrical resistivity, ρ and conductivity, σ as a function of magnetic field, H is plotted in Fig. 4(a) and 4(b), respectively. As seen in Fig. 4(a), the resistivity decreases nearly linear with increasing temperature in the range of 50–300 K. However, this behavior is distinct from that of conventional semiconductors, which typically exhibit an exponential decrease in resistivity with temperature [11]. The resistivity at 300 K is measured to be approximately 1.8 mΩ·cm, which is comparable to that of CoFeCrAl (0.8 mΩ·cm), SGS [32]. The linear decrease in resistivity with temperature corresponds to a negative temperature coefficient of resistivity (TCR) of approximately $-3.3 \times 10^{-6}$ Ω·cm/K, which falls within the typical range observed in other spin gapless semiconductors such as $Mn_2CoAl$ [11], CoFeCrGa [33], and CoFeCrAl [32]. These observations indicate that CoRuTiGe demonstrates spin gapless semiconductor–like transport behavior.

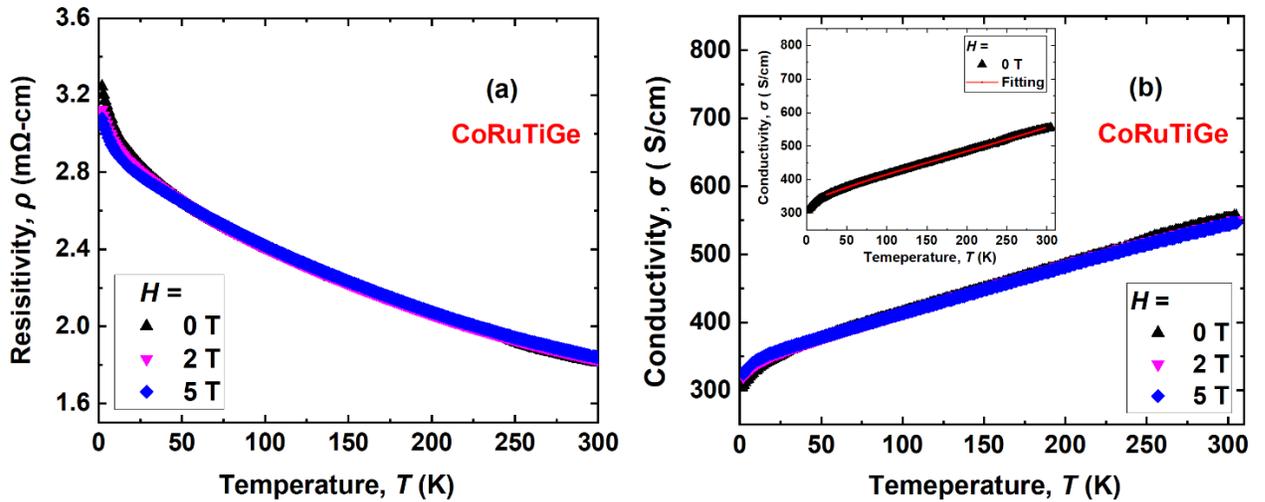

**FIG. 4.** (a) Temperature, $T$ dependence of (a) electrical resistivity, $\rho$ and (b) electrical conductivity, $\sigma$ at 0, 2 and 5T fields for CoRuTiGe. The Inset in (a) shows the linear fitting to resistivity and (b) shows the fitting of conductivity data using equation (3).

To understand the electrical transport properties of CoRuTiGe, the conductivity data was analyzed using the two-carrier model proposed by Kharel *et al.* [32]. This model has been successfully used to describe the conductivity behavior of various narrow-gap, gapless semiconductors as well as disordered SGSs. The conductivity was fitted using an equation that includes contributions from both the gapless spin channel ($\sigma_{SGS}$) and the semiconducting channel ($\sigma_{SC}$).

$$\sigma(T) = \sigma_{SGS} + \sigma_{SC} \qquad (2)$$

The above equation can be re-written as [30]:

$$\sigma(T) = \sigma_o \left[1 + 2ln(2)\frac{K_B T}{g}\right] + \sigma_{SC}\, exp\left(\frac{-\Delta E}{K_B T}\right) \qquad (3)$$

Here $\sigma_o$ is the gapless conductivity at 0 K, $\sigma_{SC}$ is the semiconducting contribution, $K_B$ is the Boltzmann's constant, $\Delta E$ is activation energy, and $g$ is a parameter reflecting the overlapping of bands. Equation (3) was fitted to experimental conductivity data and the parameters obtained from the fitting are found to be $g$ = 46 meV, and $\Delta E$ = 26.9 meV (Inset Fig. 4(b)). The value of $g$ indicates an overlap between the conduction and valence bands in the majority spin sub-band. This suggests a deviation from the ideal spin gapless semiconductor behavior, where no band overlap is expected. Such an approach has previously been used to analyze disordered SGS systems, where band overlap arises due to structural disorder or intrinsic defects within the material. The deviation in $g$ therefore supports the presence of disorder-induced modifications in the band structure of CoRuTiGe [32,34].

### D. Hall effect

Hall measurements for CoRuTiGe were carried out as a function of temperature and magnetic field up to 5 T, applied perpendicular to the sample plane as shown in Fig. 5(a). In ferromagnetic materials, the Hall effect typically includes both ordinary and anomalous contributions. The total Hall resistivity, $\rho_{xy}$ can be expressed as [35]:

$$\rho_{xy} = R_o \mu_0 H + R_s M \qquad (4)$$

Here, the first term represents the ordinary Hall contribution, while the second term corresponds to the anomalous Hall contribution. $R_o$ and $R_s$ are the ordinary and anomalous Hall coefficients, respectively, and $\mu_0$ is permeability of free space.

The Hall conductivity, $\sigma_{xy}$ of the sample can be calculated using the following relation [36]:

$$\sigma_{xy} \cong \frac{\rho_{xy}}{\rho_{xx}^2} \qquad (5)$$

where $\rho_{xx}$ is the longitudinal resistivity. The Hall conductivity, $\sigma_{xy}$ calculated using equation (5), is shown in Fig. 5(b). The charge carrier concentration, $n$ at different temperatures was estimated by linear fitting of the Hall resistivity data at high magnetic fields and was found to be of the order of ~ $10^{21}$ cm$^{-3}$. Both the carrier concentration and mobility were observed to be nearly temperature-independent, which is a characteristic feature of spin gapless semiconductors [5] as shown in Fig. 6(a). To extract the anomalous Hall resistivity, $\rho_{xy}^A$ the y-intercept of the linear extrapolation of the high-field region of the Hall resistivity curve was used. The temperature dependence of $\rho_{xy}^A$ shown in Fig. 6(b), exhibits a trend similar to that of the zero-field longitudinal resistivity.

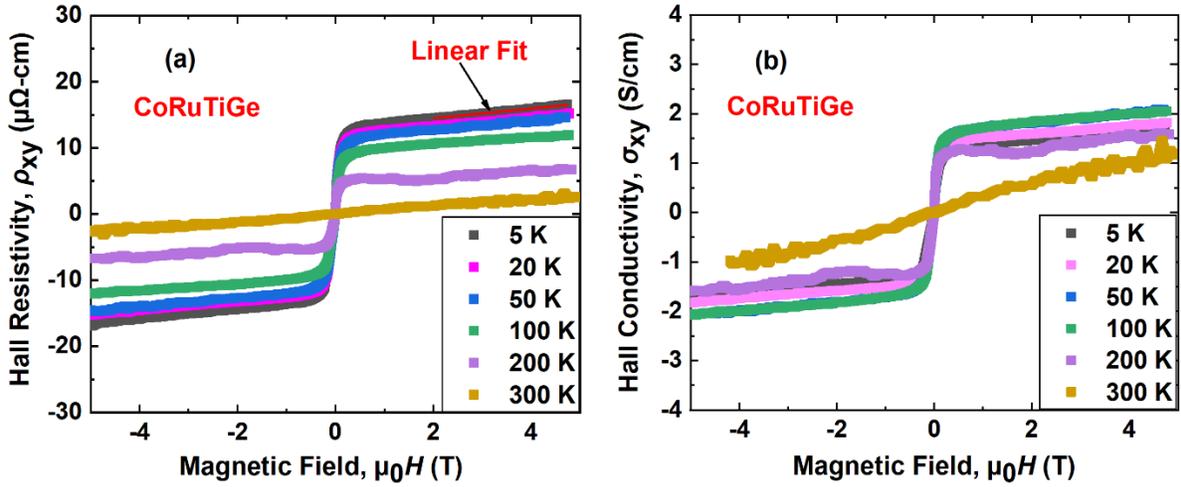

**FIG.** 5. The field dependence of (a) the Hall resistivity, $\rho_{xy}$ and (b) the Hall conductivity, $\sigma_{xy}$ as a function of temperature for CoRuTiGe.

The anomalous Hall effect (AHE) is primarily governed by two distinct mechanisms: an intrinsic mechanism, which originates from the Berry curvature of the electronic band structure, and an extrinsic mechanism, which arises from impurity scattering [37]. The extrinsic contribution can be further divided into two types of scattering processes: skew scattering and side-jump. Skew scattering is linearly proportional to longitudinal resistivity, $\rho_{xx}$ and results from asymmetric scattering of charge carriers due to impurities. In contrast, the side-jump mechanism is attributed to local distortions of the electronic wavefunction caused by impurity potentials, which generates a lateral displacement and thus a localized current density. Differentiating between the side-jump and intrinsic (Berry curvature) contributions is challenging, as both scales with $\rho_{xx}^2$. The overall scaling behavior of the anomalous Hall resistivity can be described by the empirical relation [38]:

$$\rho_{xy}^A = a\rho_0 + b(\rho_{xx} - \rho_0) + c\rho_{xx}^2 \qquad (6)$$

where $\rho_0$ is the residual resistivity. The coefficients $a$ and $b$ correspond to skew scattering contributions from defect-induced and phonon-induced scattering, respectively, while $c$ accounts for the combined intrinsic (Berry curvature) and side-jump effects. Fig. 6(c) shows the fitting of $\rho_{xy}^A$ using equation (6), yielding the of parameters: $a = 0.041$, $b = 0.068$ and $c = -1.116 \times 10^{-5}$ $(\mu\Omega \cdot cm)^{-1}$. The results clearly indicate that the AHE in CoRuTiGe arises from a combination of intrinsic and extrinsic mechanisms. The value of $b$ is found to be larger than $a$, suggesting that phonon-induced skew scattering is the dominant extrinsic mechanism. Additionally, the small value of $c$ indicates a minor contribution from Berry curvature, i.e., the intrinsic mechanism plays a limited role in this sample. Similar trends have been observed in other materials such as CoFeMnSn [39] CoFeMnSi film [40], and $Mn_2CoAl$ film [41].

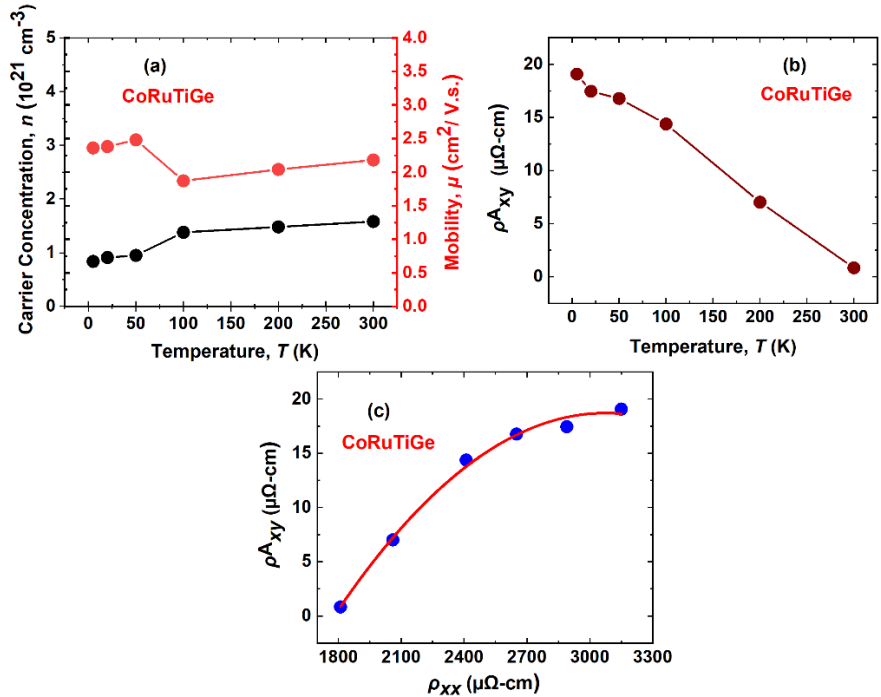

**FIG.** 6. Temperature dependance of (a) charge carrier concentrations, $n$ and (b) anomalous Hall resistivity, $\rho_{xy}^A$. (c) anomalous all resistivity, $\rho_{xy}^A$ vs. longitudinal resistivity, $\rho_{xx}$ and the solid red line shows the fitting of data by equation (6).

### E. Magnetoresistance

Magnetoresistance (MR) refers to the change in a material's electrical resistance in response to an external magnetic field and is calculated using the following equation: [42]

$$\text{MR \%} = \frac{R(T,H) - R(T,0)}{R(T,0)} \times 100 \qquad (10)$$

where $R$ is an electrical resistance. Magnetic field up to $\pm 5$ T was used to record the change in resistance at 5, 20, 100, and 300 K temperatures as shown in Fig. 7. At low temperatures (5 and 20 K), the material exhibits a non-saturating, symmetric negative MR of up to 5%. In ferromagnetic materials, such negative MR typically arises from the suppression of spin-dependent scattering with increasing magnetic field [43]. However, at higher temperatures, the change in resistance with applied magnetic field becomes negligible.

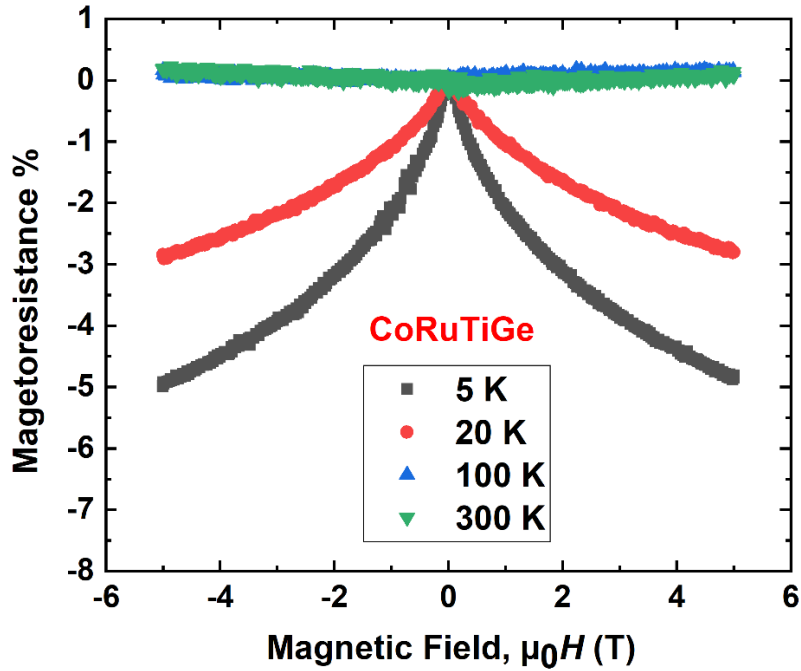

**FIG**. 7. The field dependence of magnetoresistance, MR at 5, 20, 100 and 300 K temperature for CoRuTiGe.

### F. Theoretical results
#### 1. Crystal structure and stability

In quaternary Heusler alloys with space group $F\bar{4}3m$, there are four atomic sites, $A$ (0, 0, 0), $B$ (0.5, 0.5, 0.5), $C$ (0.25, 0.25, 0.25), $D$ (0.75, 0.75, 0.75). We considered three different types of configurations as described in Table-1 and compared their total energies with respect to the Type-1. Notably Type-1 configuration is the most stable, and therefore, the Type-2, and Type-3 configurations are excluded from further discussion. The optimized lattice parameter of the Type-1 configuration is 5.986 Å, which is consistent with earlier report based on first-principles investigations [31].

**Table 1:** Considered configurations of the atomic arrangement in CoRuTiGe.

| Configuration | A (0, 0, 0) | B (0.5, 0.5, 0.5) | C (0.25, 0.25, 0.25) | D (0.75, 0.75, 0.75) | Energy (eV/f.u.) |
|---|---|---|---|---|---|
| Type-1 | Co | Ru | Ti | Ge | 0 |
| Type-2 | Co | Ti | Ru | Ge | 1.678 |
| Type-3 | Co | Ge | Ru | Ti | 1.133 |

As there has not been any prior report on the experimental synthesis of CoRuTiGe. So, it is important to discuss the stability of CoRuTiGe in terms of the formation energy, $\Delta E$ and the phase separation energy, $\delta E$. Negative $\Delta E$ stands for the stability of an alloy against decomposition into bulk phases of the constituent elements, whereas negative $\delta E$ indicates that the alloy is not likely to segregate to form any other secondary alloys. Formation energy of CoRuTiGe is $-0.57$ eV/atom. The phase separation energy is $+0.08$ eV/atom. Note that we considered ternary Heusler alloys, such as $Co_2TiGe$ and $Ru_2TiGe$ as the competing phases. Earlier studies support that the alloys with marginally positive phase separation energy ($< 0.1$ eV/atom) is synthesizable in experiment [44,45].

## 2. Electronic structure of ordered phase

Fig. 8 presents the spin-resolved band dispersion and density-of-states (DOS) for CoRuTiGe in its perfectly ordered, $Y$ structure. It has a half-metallic electronic structure owing to the presence of a band gap of 0.35 eV in the minority spin channel. On the other hand, the top of the valence band just touches the Fermi level, resulting in a pseudo gap and a very small DOS in the majority spin channel at the Fermi level. Thus, CoRuTiGe can be considered as a nearly spin-gapless semiconductor in its ordered structure. The scenario is quite similar to our earlier investigation on the isoelectronic, quaternary Heusler alloy, CoRuTiSn [27].

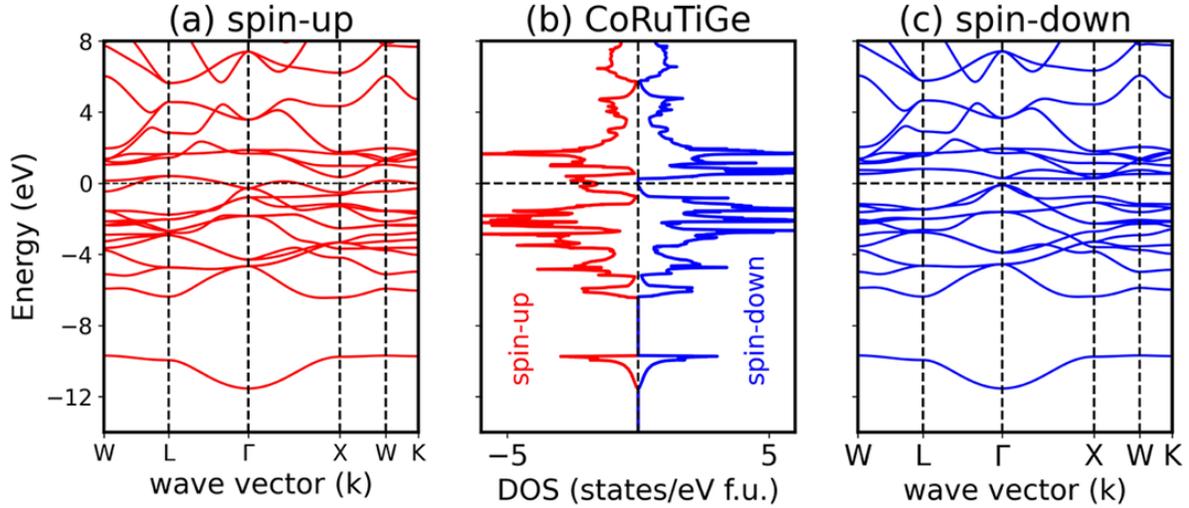

**FIG.** 8. Spin-resolved band structure and density-of-states (DOS) of ordered CoRuTiGe.

### 3. Magnetic moment and Heisenberg exchange coupling constant

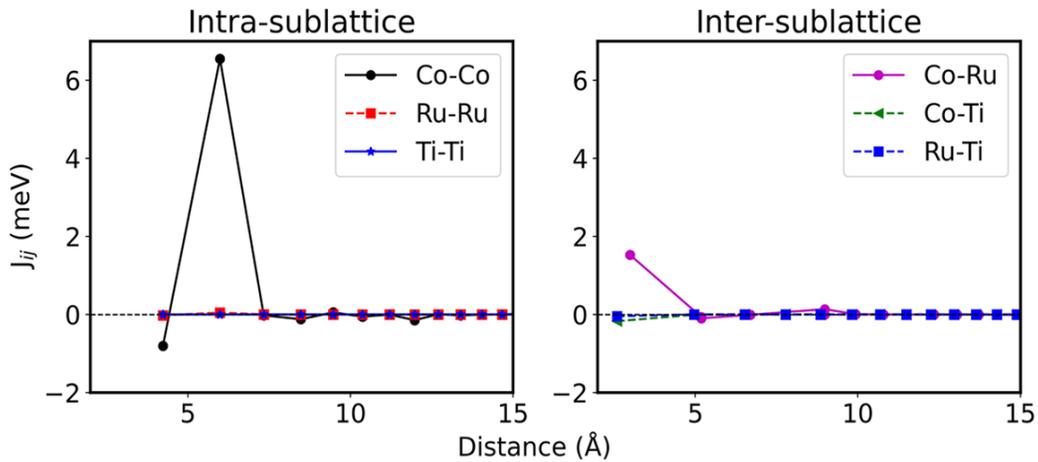

**FIG**. 9. Heisenberg exchange coupling parameters as a function of interatomic distance.

Calculated total magnetic moment of CoRuTiGe is 1.00 $\mu_B$, which satisfies the Slater-Pauling rule [46]. Co, Ru, Ti, and Ge atoms carry a magnetic moment of 0.86 $\mu_B$, 0.21 $\mu_B$, -0.03 µB, 0 µB, respectively. Now to understand the magnetic interactions between the constituent elements, we present Heisenberg exchange coupling constants ($J_{ij}$) between the $i^{th}$ and $j^{th}$ atoms as a function of the interatomic distance ($R_{ij}$) in Fig. 9 for the ordered phase. The positive (negative) values of $J_{ij}$ represent ferromagnetic (antiferromagnetic) interactions between $i^{th}$ and $j^{th}$ atoms. Noting that Co is the element with the largest magnetic moment in CoRuTiGe, the strongest interaction is between Co atoms themselves, which is a long range oscillatory RKKY type of interaction. Although, the stronger interaction between Co atoms is limited up to second nearest neighbor or the boundary 1$^{st}$ unit cell (~ 6 Å), however, a weak Co-Co interaction is extended beyond the second unit cell (~ 12 Å). Other magnetic interactions are between Co and Ru, Co and Ti, and Ru and Ti atoms. These interactions are reasonably weak. Thus, the spin directions of Ru, and Ti atoms could be more susceptible and eventually flip due to the presence of external perturbations, such as thermal excitation, tetragonal distortion of the unit cell, which could also be a reason behind the smaller magnetic moment from experiment 0.68 µB. We also evaluate the Curie temperature based on the mean-field-approximation, which is about 230 K, which is close to experimentally observed value.

## 4. Stability of swapping disorder and its impact on electronic structure

In the experimental synthesis of quaternary Heusler alloys, it is indeed a challenge to fabricate ordered phase or *Y* type structure, because of possible antistites disorders. These antistites disorders could impact on the electronic structure as well as on the magnetism. In this section we discuss the stability of the formation of disorders resulting from site exchange of the constituent atoms. Fig. 10 shows a comparison of the total energies and magnetic moments after considering a number of possible site exchanges, such as between (i) Co and Ru (*L*2$_1$-I), (ii) Ti and Ge (*L*2$_1$-II), (iii) Co and Ru, and Ti and Ge, simultaneously (*B*2), (iv) Ru and Ti atoms (*XA*).

We find that in all the cases except *XA*-type, the total magnetic moment remains very close to 1.00 µB, thus not sensitive against the disorders. However, in *XA*-type, the total magnetic moment becomes 1.45 µB. Now, we compare the relative stability of the respective disordered phases with respect to the ordered one in terms of total energies. We find that *L*2$_1$-I (-0.093 eV/f.u.) is the energetically most favorable, followed by and *Y* (0 eV/f.u.), *B*2 (+0.380 eV/f.u), *L*2$_1$-II (+ 0.422 eV/f.u.) types structures. It is to note that *XA* type (+1.411 eV) is energetically the most unfavorable, thus least likely to be synthesized in experiment.

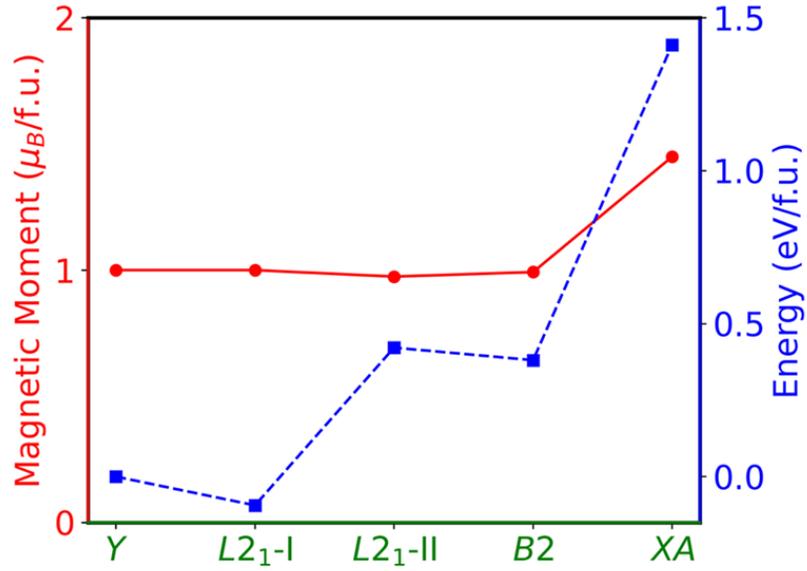

**FIG.** 10. Total magnetic moment and total energy for the ordered and disordered structures.

Fig. 11 shows the impact of different swapping disorders on the DOS. In each case, the DOS of the disordered one (Red solid line) is compared with that of the ordered one (broken blue line). We find that the DOS around the Fermi level is least affected by the swapping of Co and Ru atoms, namely $L2_1$-I, type of disorder. Half-metallicity is maintained in all types of disorders except *XA*-type, which involves swapping of Ru and Ti atoms.

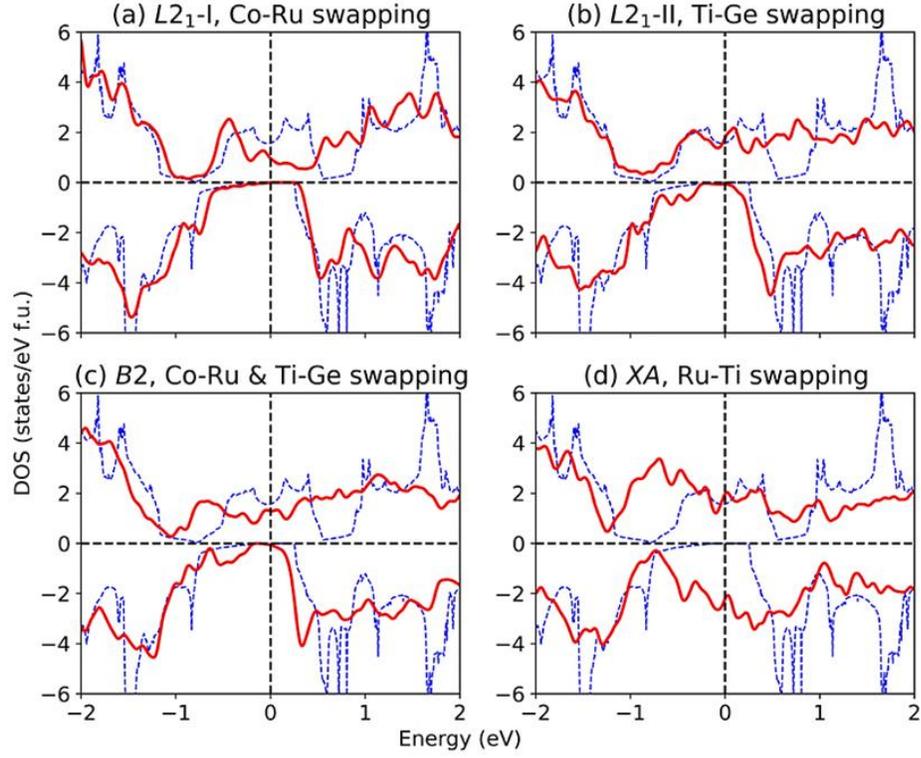

**FIG.** 11. Solid red lines show the density-of-states for the disordered phases. Dashed blue line shows the density-of-states for the ordered phase.

### 5. Anomalous Hall conductivity

In Fig. 12(a), we present the anomalous Hall conductivity (AHC), $\sigma_{xy}$, as a function of the Fermi energy position. We find that at Fermi energy, $\sigma_{xy} = 79$ S/cm which is much larger than the experimental value. Note that the calculational value is obtained for the ordered structure of $Y$ type. We could also observe a peak value of $\sigma_{xy} = 150$ S/cm at about 0.2 eV below the $E_F$. This peak value corresponds to band crossing between spin-up and spin-down bands near the Γ point as enclosed by the rectangular box in Fig. 12(b). We also present the zoomed-in view of the rectangular box in presence and in the absence of SOC as marked by the arrows. We find that the consideration of SOC leads to band anti-crossings at the points where spin-up and spin-down bands intersect each other.

These band anti-crossings act as the source of Berry curvature and of intrinsic AHC in the presence of SOC [46]. Thus, we believe successful fabrication of ordered structure, along with the tuning of the Fermi level position at about -0.2 eV in terms of the hole doping could significantly enhance the intrinsic contribution to the AHC.

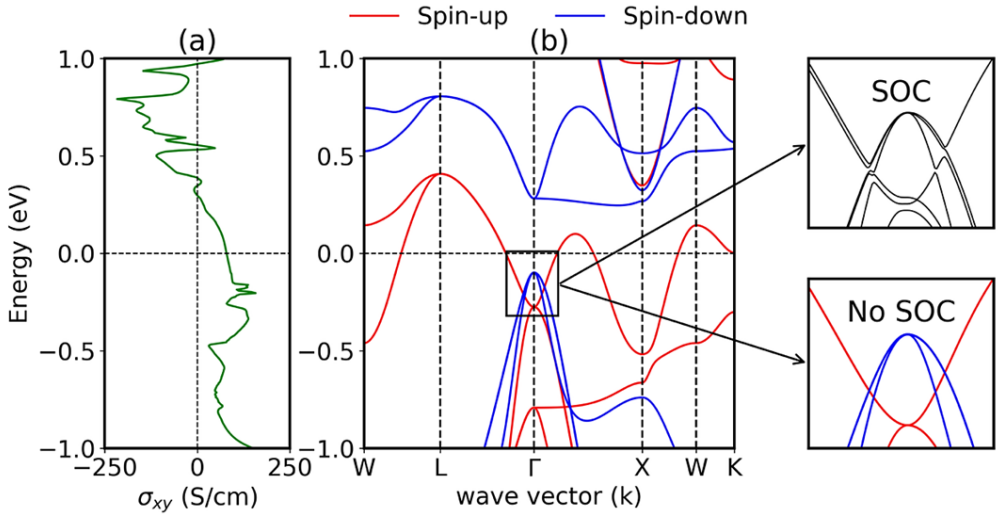

**FIG.** 12. (a) Anomalous Hall conductivity as a function of Fermi energy. (b) Spin-resolved band dispersion. Insets show zoomed-in views of the rectangular box marked in (b). The top inset shows the band dispersion in the presence of SOC, while the bottom inset in the absence of SOC.

## IV. Conclusions

The quaternary Heusler alloy CoRuTiGe was synthesized using arc melt technique. The crystal structure from XRD pattern was found to be tetragonal. The material shows ferromagnetic to paramagnetic magnetic transition with negligible hysteresis and saturation magnetization of ~ 0.681 $\mu_B$/f.u. at 5 K. The resistivity decreases linearly with temperature, and the temperature-independent carrier concentration and mobility suggest spin gapless semiconductor–like behavior of CoRuTiGe. It shows negative MR with the field at low temperatures, which becomes negligible at room temperature. SGS like properties of CoRuTiGe makes it a promising material for spintronic applications.

**Acknowledgments**

The authors sincerely thank Prof. K. G. Suresh, IIT Bombay, for experimental support in material synthesis and Dr. Amitabh Das for insightful discussions. S.G. gratefully acknowledges the financial support from the Anusandhan National Research Foundation (ANRF), New Delhi, under project SUR/2022/004713 for facilitating this research work. B.G. would like to acknowledge Shiv Nadar Institution of Eminence, Delhi NCR for the fellowship. This work was in part supported by the Initiative to Establish Next-Generation Novel Integrated Circuit Centers (X-NICS) (Grant No. JPJ011438) from Ministry of Education, Culture, Sports, Science and Technology (MEXT), and by Center for Science and Innovation in Spintronics (CSIS), Tohoku University.